# Giant supermagnonic Bloch point velocities by jet propulsion effect in cylindrical ferromagnetic nanowires

F. Tejo[1*], J.A. Fernandez-Roldan[2,3], K.Y. Guslienko[4,5], R. Otxoa[6,7] and O. Chubykalo-Fesenko[8*]

[1] Escuela de Ingeniería, Universidad Central de Chile, Santiago de Chile, 8330601, Chile.
[2]Institute of Ion Beam Physics and Materials Research, Helmholtz-Zentrum Dresden-Rossendorf e.V., Bautzner Landstrasse 400, Dresden, 01328, Germany.
[3]Department of Physics, University of Oviedo, Oviedo, 33007, Spain.
[4]Depto. Polímeros y Materiales Avanzados: Física, Química y Tecnología, Universidad del País Vasco, UPV/EHU, San Sebastián, 20018, Spain.
[5]IKERBASQUE, the Basque Foundation for Science, Bilbao, 48009, Spain.
[6]Hitachi Cambridge Laboratory, J. J. Thomson Avenue, Cambridge, CB3 0HE, United Kingdom.
[7] Donostia International Physics Center, San Sebastian, 20018, Spain.
[8]Instituto de Ciencias de Materiales de Madrid, 28049 Madrid, Spain, CSIC, Sor Juana Inés de la Cruz, 3, Madrid, 28049, Spain.

*Corresponding author(s). E-mail(s): felipe.tejo@ucentral.cl; oksana@icmm.csic.es;
Contributing authors: j.fernandez-roldan@hzdr.de; kostyantyn.gusliyenko@ehu.eus; ro274@cam.ac.uk;

**Abstract**

Achieving high velocities of magnetic domain walls is a crucial factor for their use as information carriers in modern nanoelectronic applications. In nanomagnetism and spintronics, these velocities are often limited either by internal domain wall instabilities, known as the Walker break- down phenomenon, or by spin wave emission, known as the magnonic regime. In the rigid domain wall model, the maximum magnon velocity acts as an effective "speed of light", providing a relativistic analogy for the domain wall speed limitation. Cylindrical magnetic nanowires are an example of systems with the absence of the Walker breakdown phenomenon. Here we demonstrate that in cylindrical nanowires with high magnetization such as Iron, also the magnonic limit could be out- standingly surpassed. Our numerical modelling shows the Bloch point domain wall velocities as high as 14 km/s, well above the magnonic limit estimated in the interval 1.7-2.0 km/s. The key ingredient is the conical shape of the domain wall which elongates and breaks during the dynamics, leading to domain wall acceleration due to the jet propul- sion effect. This effect will be very important for three-dimensional spintronic networks based on cylindrical magnetic nanowires.

# Introduction

Domain walls (DWs) are topological magnetic solitons with a huge area of potential applications in modern nanoelectronics and spintronics. Magnetic domain walls are currently used for many applications such as the race-track memory in information technologies [1, 2], logical circuits[3], neuromorphic computing[4], etc. Apart from technological perspectives, there is a fundamental question of the key mechanism limiting propagation of any travelling wave in a given media. Generally, the maximum speed of information transmission cannot surpass the maximum group velocity of elementary excitations. Thus, this maximum velocity, defined as an effective "speed of light" can be viewed as an analogous to limiting velocity in the special theory of relativity [5]. Additionally, topological solitons propagation is limited by wave emission (e.g., acoustic or electro-magnetic), similar to the Cherenkov effect [6, 7] ("speed of sound"). The question arises how solid are the above restrictions and if even higher velocities can be achieved.

Achieving high domain wall velocities is a key ingredient for many applications because they determine the device operation time. Unfortunately, in ferromagnetic materials the domain wall velocity also suffers from the so-called Walker breakdown phenomenon[8]. In Permalloy (FeNi alloy) stripes (widely used for spintronic and magnonic studies), DW velocity cannot exceed hundreds of m/s [9]. Consequently, a lot of research has been performed with the aim to find solutions of this problem. The manipulation of ferromagnetic DWs with spin-orbit torques in insulator/metal heterostrictures with the Dzyaloshinskii-Moriya exchange interactions and out-of-plane magnetization resulted in DW velocities up to 400 m/s [10]. Differently to the above materials, ferrimagnetic and antiferromagnetic materials do not have the Walker breakdown phenomena. In ferrimagnetic garnets the DW velocities up to 4.3 km/s have been recently reported [5]. Also, in ferrimagnetic materials for temperatures around a net angular momentum compensation point, high velocites up to 2 km/s [11] have been measured. In pure antiferromagnets with special crystal lattice symmetries such as $Mn_2Au$ even higher velocities up to 40 km/s were theoretically predicted [12–15]. The DW speed in these materials is limited by the magnon group velocity [13], used as an analog to the "speed of light" for their dynamics which obey equations similar to ones in the spe-cial theory of relativity. After reaching the corresponding velocity, the DW is slowed down and a proliferation of multiple domain walls, arising from the initial wall has been observed in numerical modelling [13].

Cylindrical magnetic nanowires are viewed as building blocks for three-dimensional applications such as internet of nano things [16]. They constitute another example of systems where the absence of the Walker breakdown phenomenon has been predicted [17]. Starting with some diameters, cylindrical magnetic nanowires possess a very special DW of the Bloch point type (BP-DW) [17–19], in which head-to-head or tail-to-tail magnetic domains are separated by a vortex domain wall containing a Bloch point (see Fig. 1). The Bloch point (BP) is a 3D magnetic topological soliton (hopfion), whose configuration has a singularity of the magnetization field in its center with an excess of the exchange energy. It was predicted that BP has a large mobility in response to external driving magnetic field [20]. Recent numerical simula- tions reported the BP-DW velocity in Permalloy cylindrical nanowire circa 1 km/s, slightly higher than the magnon group velocity [21]. After that, the DW dynamics are entering in a so-called turbulent regime characterised by prolifer-ation of magnetic "drops"[22]. Unfortunately, the dynamics of BP-DWs have another problem, since it has been shown that they undergo dynamical trans- formation to the vortex-antivortex domain wall [23, 24]. This phenomenon would have a large impact and may limit possible applications of cylindrical magnetic nanowires[23].

In this article, we show that the BP-DW velocity in ferromagnetic cylindri- cal nanowires of high magnetization (such as Iron) can reach very high values after overcoming the turbulent magnonic regime, i.e., in the "super-sonic" regime. Fig. 1 illustrates the internal magnetization texture of the BP-DW. Being initially of a symmetrical shape at rest (Fig.1a), as it moves along the nanowire under an applied field, the BP-DW develops a conical shape in which the BP is positioned at the vertex of the cone (Fig.1b). The length of the cone increases as the time progresses. As we show later, some internal instabilities break the cone-shape DW, creating a pair of new BPs. This produces a jet propulsion effect, accelerating the original BP. Our numerical modelling shows the BP velocity values up to 14 km/s, much higher than any value reported previously in numerical modelling of ferromagnetic nanowires.

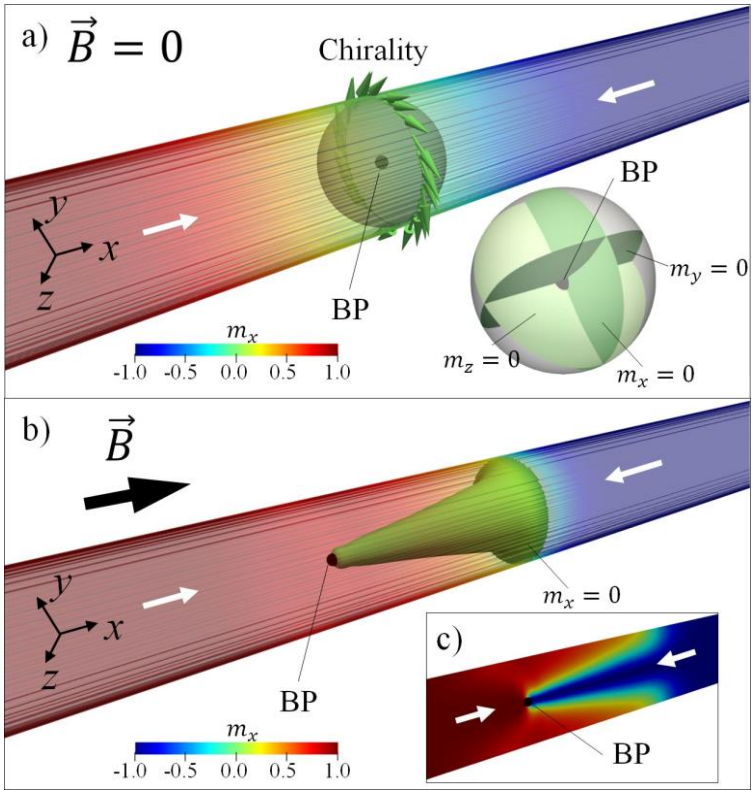

**Fig. 1** a) The magnetization configuration of the cylindrical nanowire in the absence of magnetic field. Arrows on the wire surface represent the direction of magnetization in the domain wall plane. The location of the Bloch-point is defined as the position of the intercession of the iso-surfaces where all three magnetization components $m_x = m_y = m_z = 0$. b) Snapshot of magnetization in the presence of a magnetic field applied in the direction of the wire axis. The iso-surface $m_x = 0$ forms a cone at the vertex of which the BP is located.

## Results

As an illustrative example, we first discuss the dynamics of a BP-DW in a nanowire with the radius $R = 30$ nm, length $L = 1.5$ $\mu$m (more details of the model could be found in the Methods section). An important parameter to describe BP-DW is its chirality, i.e., direction of the magnetization circulation in the yz plane perpendicular to the nanowire x-axis. We start our simulations with DW of "good" chirality, as defined by Ref.[17], i.e., the one with a circular polarisation direction favored by the torque induced by internal field (depicted in Fig. 1). An external magnetic field *B* is applied parallel to the nanowire axis. Fig. 2a shows an "effective" DW position as a function of time, evaluated from the zero value of the average magnetization along the nanowire axis, i.e., $<m_x(t)> = 0$ (an experimentally measured quantity) for different driving magnetic field magnitudes. Fig. 2b shows the BP position along the nanowire axis for the same driving fields. The BP position has been calculated by intersecting three iso-surfaces corresponding to zero values of the magnetization components $m_x = 0$, $m_y = 0$, $m_z = 0$. Importantly, both Figures show similar tendencies with the exception of the void region in Fig. 2b, which corresponds to the absence of the BP in simulations. Fig. 2 presents three regimes for the BP-DW motion. First, for applied fields *B* <15 mT the DW propagation is approximately linear in time and DW velocity increases as the magnitude of external field increases. Secondly, for applied fields 15 mT < *B* < 60 mT the DW mobility is significantly decreased. This regime is characterized by a large emission of spin waves from the BP-DW towards the nanowire edges which was associated previously with the spin-Cherenkov effect[7, 17, 21]. Note that the Cherenkov radiation starts when the moving DW reaches the minimal phase velocity of the linear spin waves and exceeds it [7]. Also, the DW texture undergoes a series of complex transformations. Moreover, for *B* = 35 mT during some time interval the BP does not exist. These effects are similar to previous works on transformation of BP-DW into vortex-antivortex DW [23]. This regime has been reported previously as the one limiting BP-DW propagation [23]. Additional illustrations can be found in the supplementary information. The novelty comes in the third regime for *B* > 60 mT where we again discover a cuasi-linear DW motion with time with a very high velocity. This is a regime at which we will mostly put our attention.

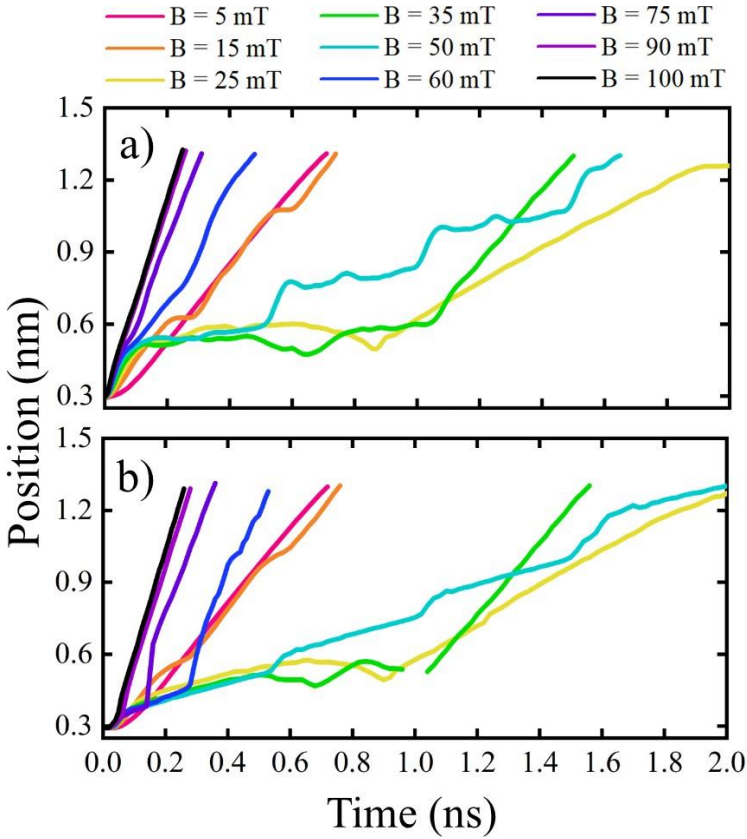

**Fig. 2** a) Effective Domain Wall position evaluated from the dynamics of the longitudinal magnetization component $<m_x>(t)$, as its zero crossing point and b) Bloch-Point position as a function of time. The calculations correspond to a nanowire with 30 nm radius and different magnetic field magnitudes.

Figure 3 summarizes the maximum BP velocity as a function of applied magnetic field for two different nanowire radii and two damping parameter values (note that in the initial interval of time, see, e.g. $B = 60$ mT and $B = 75$ mT in Fig. 3b, the BP mobility is low and we extract the velocity from the high mobility region). We observe that the critical magnetic fields limiting the transition between different regimes displace to higher values as the nanowire radius decreases. As an example, for $R$ = 20 nm (see Supplementary Material) we observed only the first and the second DW motion regimes within the same field interval as considered in Fig. 3, while for $R$ = 40 nm the first regime is displaced to small fields. This transition between the first and the second regime occurs in all cases for velocities of approximately 1.7-2.0 km/s, which we associate with the Cherenkov effect. Theoretical estimation of the minimum phase magnon velocity, determining the start of the magnon emission [7] can be found in our Supplementary Material. Its value is in the interval 1.5-1.8 km/s, it decreases as a function of nanowire radius and slightly increases as a function of applied field. In simulations, approaching this velocity produces a large spin wave emission and either a stagnation or a decrease of the DW velocity. However, in the third regime the velocity are significantly increased up to the value of circa 4 km/s for nanowires with 30 nm radius and 14 km/s for 40 nm radius. We notice also the increase of velocities as the nanowire radius increases or the damping value decreases.

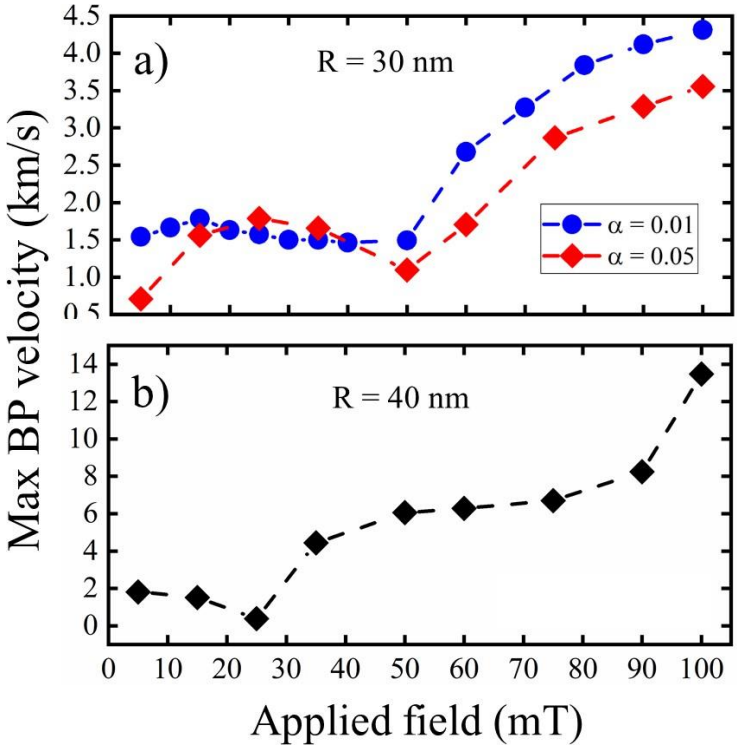

**Fig. 3** Maximum Bloch-point velocity as a function of applied magnetic field for nanowires with radii a) 30 nm and two damping parameter values, $\alpha$ and b) 40 nm and $\alpha$ = 0.01. For all results, a cell size of $1\times1\times1$ nm$^3$ was used in the simulations.

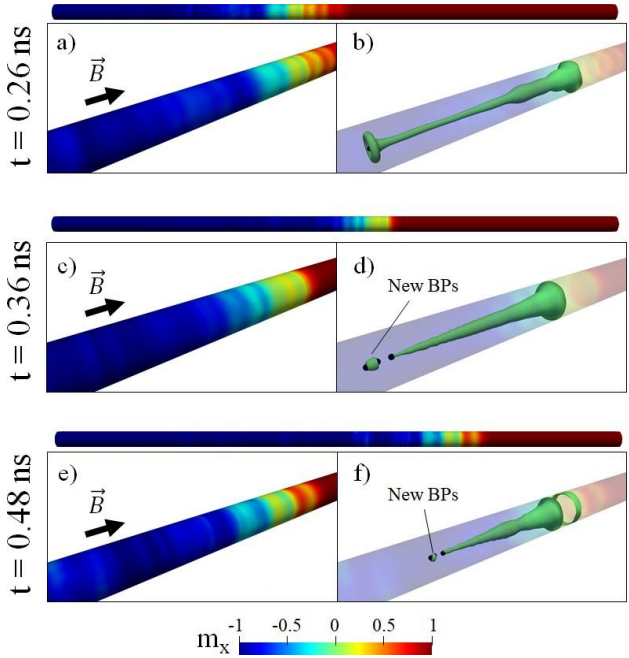

**Fig. 4** Snapshots of the magnetization dynamics when a magnetic field of 60 mT is applied. a), c) and e) show the surface magnetization of the nanowire at different times, while b), d) and f) show the internal magnetization configuration of the domain wall. The greencone-shaped surface represents an iso-surface where the *x*-component of the magnetization is zero, and the black points indicate the position of the BPs.

To clarify the origin of these behaviors, we show in Fig. 4 snapshots of the magnetization dynamics for $B$ = 60 mT. Since we are above the effective "speed of sound", the emission of spin waves is very relevant, especially visible on the nanowire surface magnetization. The spin waves are constantly being emitted, mostly in front of the BP on the nanowire surface but also in the backward direction. In the domain wall region, they may be considered as part of the domain wall, forming a cloud around and propagating together with it (see available Supplementary Video). However, the conical structure of the BP-DW is conserved, and its propagation is not interrupted. Importantly, during the dynamics, the length of the cone increases and the cone breaks at some instant of time, giving rise to a spontaneous creation of a pair of BPs near the cone end. This pair is formed by a head-to-head and a tail-to-tail BPs with opposite chiralities. The original (head-to-head) BP detaches from the cone together with one of the new born (tail-to-tail) BPs and the other new BP (identical to the original one) is now located at the DW cone end. Additional illustration can be found in the Supplementary Material. After some time, the detached part of the cone BPs annihilates leading to further spin wave emission. In our simulations, we always follow the dynamics of the BP which is attached to the main cone, i.e., forms part of the DW. This mechanism of creating additional BPs provides a propulsion effect on the BP-DW, leading to a change in the slope for the DW or BP positions in Fig.2 and allowing the BP to reach high velocities.

Several effects are important here. First, the main contribution to the increase of the cone length comes from the magnetostatic interactions. Indeed, due to the difference of magnetostatic fields at the surface and the center of the nanowire, surface magnetization propagates faster and the one in the center lags behind, producing an elongation of the DW cone. This effect is larger in nanowires with high saturation magnetization such as Fe as well as it is larger in nanowires with larger diameters.

The next important ingredient of the processes is the fact that the cone breaks due to dynamical instabilities. This effect was noticed previously in Refs. [22] as a drop generation behind DW, however its significance for the DW velocity was not discussed and the large velocity increase was not reported. For better understanding, we systematically study the dynamics of the so-called kinetic field, used in the studies of dynamical instabilities of magnetic solitons such as reversal of the core polarity in magnetic vortices by vortex-antivortex pair nucleation [25] or DW-pairs proliferation in layered AFM such $Mn_2Au$[13]. The kinetic field is defined via the gauge vector potential introduced to keep the Lagrangian invariant [26] and is written as:

$$\boldsymbol{h}_{kin} = \frac{1}{M_s} \frac{\partial L_{kin}}{\partial \boldsymbol{m}},$$

where

$$L_{kin} = \frac{M_s}{\gamma} \frac{(\boldsymbol{m} \times \dot{\boldsymbol{m}}) \boldsymbol{n}}{1 + \boldsymbol{m} \cdot \boldsymbol{n}},$$

corresponds to the kinetic part of the Lagrangian of our system, where **m** is a unit vector in the direction of magnetization and **n** is a unit vector of Cartesian coordinates. Figures 5a and 5b compare the maximum kinetic field magnitude and the number of existing BPs along the BP dynamical trajectory. Importantly, new BPs are constantly generated and annihilated (in pairs and with opposite chiralities) and in this case up to five BPs can exist in the system simultaneously. One can associate the appearance of additional BPs with a burst in the kinetic field. The increase of the DW velocity can be also associated with the start of this process, i.e., the DW velocity was low when only one BP was present in the system and the DW accelerates due to the appearance of the new born BP-pair. The rate of additional BP production increases with the driving field increasing. For higher fields (see Supplementary Material for $B$ = 100 mT) the process of creation/annihilation of the BP pairs occurs continuously during the propagation of the BP-DW and it takes place very close to the BP that is part of the DW, allowing the BP-DW to achieve a continuous displacement at high velocity and additional acceleration, comparatively to the $B$ = 60 mT case, where the velocities boosts are visible in the DW trajectory.

Finally, the acceleration of the BP, attached to the DW main cone, is due to the propulsion effect. Indeed, by breaking the cone, the cone size substantially decreases. After the breaking, the detached part carries a linear momentum and accelerates the dynamics of the main cone. The underlying mechanism in

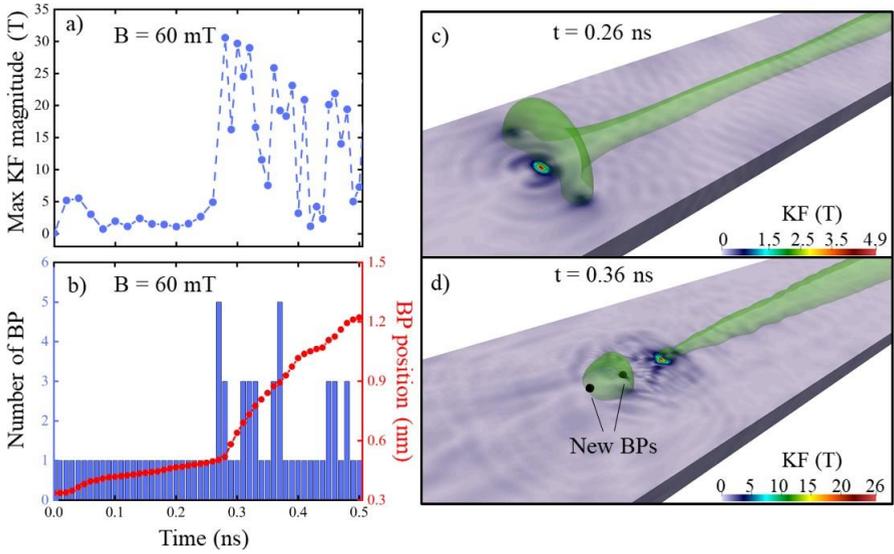

**Fig. 5** a) Maximum magnitude of the kinetic field as a function of time. b) Amount of BPs in the nanowire as a function of time. c) and d) Spatial representation of the kinetic field through a cross section of the nanowire at two different time instants.

this case is related more to the exchange interactions than to the magnetostatic ones. Indeed, the BP in the cone has an opposite chirality to the one of the newborn, closest to it, and the exchange interaction produces a repulsion effect.

## Discussion

Cylindrical geometry offers possibilities to achieve very high velocities of magnetic domain walls due to the absence of the Walker breakdown phenomenon. However, because of the possible dynamical instabilities of the BP-DWs and their transformation to other magnetization configurations, this possibility has been questioned [23]. In our numerical modelling, we report very high veloc- ity (up to 14 km/s) of the BP-DW in cylindrical magnetic nanowires of high magnetization value such as Fe. These velocities are much higher than the "magnonic limit", estimated here as being close to 2 km/s. We associate the increase of velocity to the jet-propulsion effect by a constant ejection of new BPs from the domain wall. The key to this effect is a cone shape of the DW due to different magnetization propagation speed on the nanowire surface and in its center. The effect is larger in nanowires with high saturation magnetization due to its magnetostatic origin. The internal instabilities break the cone and give birth to a pair of DWs which are ejected by the main DW. This creates a jet propulsion effect, accelerating the DW propagation along of the nanowire axis.

Our findings raise again the question about the limits for the domain wall velocities in ferromagnets. In our view, the "relativistic limit" of the domain wall velocity related to the magnon extreme velocities is valid only for the case of (almost) rigid domain wall dynamics in 1D case, while our case is very far from this. We demonstrated that magnonic limit does not restrict the velocity for Bloch point domain wall in the ferromagnetic nanowires with essentially 3D magnetization configurations and that repulsive interactions between the BPs can help to overcome it to a very large extent.

Our results will be very important for the development of spintronic devices based on three-dimensional magnetization textures since the magnetic nanowires most probably will be their parts. It is important that in this case there are only little limitations in terms of magnetic materials, for example, there is no need of materials having special crystallographic symmetry, antiferromagnetic coupling, Dzyaloshinskii-Moriya interaction or large spin-orbit torque effects.

## Methods

Simulations were performed using the GPU-accelerated micromagnetic solver MuMax3, which numerically integrates the Landau-Lifshitz-Gilbert equation of magnetization motion[27]. We considered a single nanowire with three different radii $R_i$, each associated with different lengths $L_i$, conserving approximately the wire aspect ratio $L/R$, that is: $R_1$ = 20 nm, $L_1$ = 1200 nm; $R_2$ = 30 nm, $L_2$ = 1500 nm and $R_3$ = 40 nm, $L_3$ = 2500 nm. These parameters were chosen to ensure the existence of BP-DW [19]. We used the magnetic parameters of Fe, where the saturation magnetization is $M_s$ = 1700 kA/m, and the exchange stiffness is $A_{ex}$ = 21 pJ/m. Additionally, we considered a polycrystalline system, so we have disregarded the magnetocrystalline anisotropy. Discretization size was chosen 1 nm, which produces very little effect on DW dynamics (see Supplementary Material). Unless specified we used the damping parameter value $\alpha$ = 0.01. The initial state of the system is a BP-DW positioned at one of the free ends of the nanowire (see Fig. 1a). This configuration is obtained by relaxing a forced head-to-head configuration. We have ensured that the chirality of the BP-DW is appropriate to favor its movement due to the direction of the applied magnetic field [17]. To avoid the nucleation of other domains walls at the ends of the nanowire, we eliminated the surface magnetic charges at its ends, and in this way, we imitated an infinitely long wire, which allows the BP-DW to move longitudinally along the nanowire axis.

**Supplementary Material.** In Supplementary file authors present more examples of the Bloch point domain wall dynamics as a function of applied field, nanowire diameter and discretization size. Enlarged snapshot for the additional Bloch points birth is also presented for the visualisation. Theoretical estimation of the minimum magnon phase velocity is provided. Supplementary video presents an example of time-resolved magnetization dynamics of the Bloch point domain wall in a nanowire with $R$ = 30 nm under applied field $B$ = 70 mT.

**Acknowledgments.** The authors acknowledge financial support by the grants PID2019-108075RB-C31 and PID2019-108075RB-C33 funded by Ministry of Science and Innovation of Spain MCIN/AEI/ 10.13039/501100011033. K.G. acknowledges support by IKERBASQUE (the Basque Foundation for Science). The work of K.G. was partially supported by the Norwegian Financial Mechanism 2014-2021 trough the project UMO-2020/37/K/ST3/02450.The work of F.T. was supported by ANID + Fondecyt de Postdoctorado, convocatoria 2022 + Folio 3220527. This research was partially supported by the supercomputing infrastructure of the NLHPC (ECM-02).

**Authors contributions.** F.T., J.A. F.-R. and O.C.-F. conceived the project, F.T. performed micromagnetic modelling, K.G. performed spin- wave calculations. F.T.,R.O., K.G. and O.C.-F. analysed the results and were involved in discussions and critical assessment. O.C.-F- coordinated the project. F.T. and O. C.-F. wrote the paper. All authors reviewed and contributed to the paper.